\documentclass[showpacs,preprintnumbers,amsmath,amssymb]{revtex4}
\usepackage{graphicx}
\usepackage{dcolumn}
\usepackage{bm}
\usepackage{longtable}
\newcommand{\an}{{\it and }}
\newcommand{\etal}{\it{et\,al.}}

\begin{document}
\title{Scaling approach for the time-dependent Kondo model}
\author{C. Tomaras \an S. Kehrein}
\affiliation{Arnold Sommerfeld Center for Theoretical Physics, Ludwig-Maximilians-Universit\"at M\"unchen, Theresienstr. 37, D-80333 M\"unchen, Germany}
\begin{abstract}
We present a new nonperturbative method to deal with the time-dependent quantum many-body problem, which is an extension of Wegner's flow equations to time-dependent Hamiltonians. The formalism provides a scaling procedure for the set of time-dependent interaction constants. We apply these ideas to a Kondo model with a ferromagnetic exchange coupling switched on over a time scale~$\tau$. We show that the asymptotic expectation value of the impurity spin interpolates  continuously between its quenched and adiabatic value.
\end{abstract}
\maketitle

\section{Introduction}

The time-dependent quantum many-body problem has become of particular
interest within recent years. New experimental investigations 
offer many different
possibilities to study quantum systems in the presence of external driven
degrees of freedom, such as cold atomic gases \cite{NewEntry1}, 
ultrafast spectroscopy on semiconductors
\cite{MorawetzSpec}, flux pattern formation in superconducting films
\cite{pattern} or single molecule force spectroscopy \cite{molecule}.
Many of the available theoretical methods 
evaluate the quantum dynamics of such driven systems in a certain limit. If the time-scale
of the external degrees of freedom is slow with respect to the intrinsic dynamics
of the system, it is possible to perform calculations with high
mathematical rigor by means of adiabatic perturbation theory \cite{spohn-2001,Teufel1}.
Many methods apply to time-periodic perturbations. If the driving
frequency $\omega$ exceeds the largest energy-scale of the unperturbed
system, a good method is an expansion in powers of $g/\omega$, with
$g$ the coupling between the system and the perturbation, to derive
a time independent effective Hamiltonian \cite{rahav-2003-68}. In the general time-periodic case
one often follows the Floquet approach \cite{NewEntry4}. The problem of calculating the quantum evolution
of a certain initial state for an arbitrary switching procedure is
often addressed in time-dependent Keldysh perturbation theory \cite{NewEntry11}.
However, such calculations are frequently limited to linear response, and
expansions with respect to bare couplings can produce errors which 
grow in time. Furthermore, self-consistency schemes often yield
a system of coupled non-linear integro differential equations, which
are not convenient for numerical evaluation. Pure numerical techniques
are the time-dependent density-matrix renormalization-group \cite{NewEntry5},
time-dependent density-functional theory \cite{TDDFT}, and exact
diagonalization of the operator $H-i\partial_{t}$ in discretized
form. 

Here we develop a new method to deal with the driven many-body problem,
which is a similarity renormalization scheme for time-dependent Hamiltonians
based on Wegner's flow equation method for the time-independent case 
\cite{NewEntry7,NewEntry10}. We apply our new approach
to the isotropic ferromagnetic Kondo model 
\begin{equation}
H(t)=\sum_{k,\sigma}\epsilon_{k}c_{k\sigma}^{\dagger}c_{k\sigma}+\sum_{k,k'}J(t)
\:\vec S\cdot \vec s_{kk'}
\label{defferroKondo}
\end{equation}
in a situation where the ferromagnetic exchange interaction $J(t)\leq 0$ 
is switched on continuously from zero to a nonzero
constant value~$J_0$. Here $\vec S$ is the impurity spin-1/2 degree of freedom and
$\vec s_{kk'}=\frac{1}{2}\sum_{\alpha,\beta} c^\dagger_{k,\alpha} \vec\sigma_{\alpha\beta} c_{k',\beta}$
is the conduction band electron spin density.
We choose this example for illustrating our approach 
because it is a many-body system which allows for a perturbative evaluation of the truncated RG equations, which posess a nontrivial RG flow.  

The second motivation comes from
recent calculations that provided insights in the quantum
dynamics of systems subjected to a sudden interaction quench \cite{hackl-2009-102,moeckel-2008-100}. These systems exhibit nonequilibrium dynamics 
in which the expectation values of certain observables are enhanced 
with respect to their equilibrium values. 
Specifically, we consider a time-dependent Hamiltonian
\begin{equation}
H(t)=H_0+f(t)\,H_{\rm int}
\end{equation}
where the term $H_{\rm int}$ is switched on by some protocol $f(t)$,
$f(t\rightarrow -\infty)=0$,  $f(t\rightarrow +\infty)=1$.
We study the time evolution
of the system prepared in the ground state $|\Omega_0\rangle$ of $H_0$ in the
infinite past, that is $|\Omega_0(t)\rangle$ generated by $H(t)$ with
$|\Omega_0(t\rightarrow -\infty)\rangle=|\Omega_0\rangle$.
Let $O$ be any observable that commutes with $H_0$. We 
define its {\it mismatch factor}~$\mu$
\begin{equation}
\mu\stackrel{\rm def}{=} \frac{\langle O\rangle_{\rm neq}-\langle O \rangle_0}{
\langle O\rangle_{\rm eq}-\langle O \rangle_0}
\end{equation}
Here $\langle O \rangle_0=\langle \Omega_0|O|\Omega_0\rangle$ is the noninteracting
expectation value, $\langle O \rangle_{\rm eq}=\langle \Omega|O|\Omega\rangle$
is the interacting expectation value with respect to the ground state $|\Omega\rangle$
of $H=H_0+H_{\rm int}$ and 
\begin{equation}
\langle O\rangle_{\rm neq} \stackrel{\rm def}{=} \lim_{T\rightarrow\infty} \:\frac{1}{T}\,
\int_0^T dt\: \langle \Omega_0(t)|O|\Omega_0(t)\rangle \ ,
\label{defOneq}
\end{equation}
is the time-averaged nonequilbrium expectation value. 

For adiabatic switching on of $H_{\rm int}$ we have $\mu=1$ since 
$|\Omega_0(t)\rangle$ asymptotically becomes the interacting ground state $|\Omega\rangle$. 
Remarkably, in the opposite limit of sudden switching $f(t)=\Theta(t)$ 
of a weak perturbation $H_{\rm int}$
one finds the universal value $\mu=2$ in many quantum systems. 
In fact, a mismatch factor $\mu =2$ can be 
proven rigorously for discrete weakly interacting systems \cite{moeckel3} in second
order perturbation theory. The same factor
$\mu=2$ arises in many weak-coupling continuous systems like the quantum sine-Gordon 
model for $\beta^2>8\pi$ \cite{sabio} or in the ferromagnetic Kondo model with an 
infinitesimal positive magnetic field\cite{hackl-2009,hackl-2009-102}.
In the latter model it describes the ratio of the large time asymptotic impurity spin 
normalized through its equilibrium value
\begin{equation}
\mu=\frac{\langle S_z(t\rightarrow\infty)\rangle_{\rm neq}-1/2}{\langle S_z\rangle_{\rm eq}-1/2}
\label{defmuSz}
\end{equation}
For an interaction quench in a Fermi gas in $d>1$ dimensions, the mismatch factor of the
correlation-induced corrections to the momentum distribution between nonequilibrium and 
equilibrium state turned out to be the key to understanding thermalization 
\cite{moeckel-2008-100,moeckel3}. However, in this model the universal value
$\mu=2$ is only found when averaging over an intermediate well-defined time window
in (\ref{defOneq}), which describes the prethermalized regime. 

Due to the important role of the mismatch factor~$\mu$ in characterizing the
nonequilibrium dynamics, it is natural to investigate its behavior when the
protocol $f(t)$ interpolates between adiabatic and instantaneous switching.
For the Hubbard model this has been carried out
by means of lowest order Keldysh perturbation
theory \cite{moeckel-2009}. Due to the finite switching time, the dynamics becomes
adiabatic for fast modes leading to $\mu=1$ for fast modes (large momenta). 

For the ferromagnetic Kondo problem \cite{hackl-2009,hackl-2009-102}
a similar analysis of the crossover from adiabatic to instantaneous quenches 
has not yet been established.
Since it posesses a nontrivial scaling flow, it is not clear what energy scale marks the transition from adiabatic to a sudden response of the electron system. In addition, there is no intuitive picture about the fate of the renormalized exchange couplings in presence of a time-dependent vertex. Applying our new time-dependent flow equation approach, we will show that the relevant scaling  flow is just the instantaneous one (at least for the resonant infrared couplings  which characterize the dynamics of the Kondo system at large times). 
This allows us to study the real time evolution for this
non-sudden and non-adiabatic switching procedure within a non-perturbative
scheme. Here non-perturbative means that this result cannot be
obtained from summing a finite number of diagrams. The advantage of
our renormalization scheme is to avoid the usual complicated
reordering of the perturbation series labeled by the bare coupling
$J(t)$, by solving the truncated one loop RG-equations. 

Our motivation is therefore twofold: In the first
part of this paper we introduce the new formalism and explain how
it can be used for generic time-dependent many body problems. In the second
part we apply this approach to study the quantum
dynamics of the ferromagnetic Kondo model (\ref{defferroKondo}) 
subjected to a continuously switching on of
the interaction. We calculate the mismatch factor $\mu$ from (\ref{defmuSz}) as well as the
long-time behavior of the impurity magnetization curve depending on the
switching protocol.

\section{Time-dependent flow equations}

A convenient way to solve the time-dependent many-body problem
would be the application of a unitary time-dependent transformation $U(t)$
that makes the Hamiltonian 
\begin{equation}
\mathcal{H}(t)=U(t)\,H(t)\,U^{\dagger}(t)-i\,U(t)\,\frac{dU^{\dagger}(t)}{dt}
\label{defcalH}
\end{equation}
simple enough for solving the Schroedinger equation 
\begin{equation}
i\partial_{t}|\psi(t)\rangle_{U(t)}=\mathcal{H}(t)|\psi(t)\rangle_{U(t)}
\label{timedepSchroedinger}
\end{equation}
in the new basis $|\psi(t)\rangle_{U(t)}=U(t)\,|\psi(t)\rangle$.
Now even for the simplest quantum models such
as the driven two-level system, there is no general way to find a
transformation which simplifies (\ref{timedepSchroedinger}) considerably 
and therefore this approach is not commonly used.

In this paper we will develop such a general approach based on a suitable
extension of the time-independent flow equation method (method of infinitesimal
unitary transformations) \cite{NewEntry7,NewEntry10}. 
Instead of performing the change of basis (\ref{defcalH}) in one step,
it is convenient to use infinitesimal steps by solving the differential equation
\begin{eqnarray}
\partial_{B}\mathcal{H}(B,t) & = & [\eta(B,t),\mathcal{H}(B,t)]+i\partial_{t}\eta(B,t)
\label{eq:2}
\end{eqnarray}
with a suitable antihermitean generator $\eta(B,t)$
and the initial condition posed at $B=0$: $\mathcal{H}(B=0,t)=H(t)$. Eq.~(\ref{eq:2}) becomes
equivalent with (\ref{defcalH}) for the following unitary transformation 
constructed as a $B$-ordered exponential
\begin{equation}
U(t)  =  U(B=\infty,t)=\mathcal{T}_{B}\left[e^{\int_{0}^{\infty}dB\,\eta(B,t)}\right]\ .
\label{eq:1}
\end{equation}
This approach stays conceptually close to the time-independent formulation
and becomes equivalent to it for time-independent Hamiltonians~$H$. 
Now the key question is how to choose the infinitesimal
generator $\eta(B,t)$ such that the series of Hamiltonians
$\mathcal{H}(B,t)=U(B,t)(H(t)-i\partial_{t})U^{\dagger}(B,t)$ 
eventually becomes diagonal during the flow from $B=0$ to $B=\infty$. 
The suitable choice is to add an additional time derivative
to Wegner's canonical generator for the time-independent case \cite{NewEntry7}:
\begin{equation}
\eta(B,t) =  [\mathcal{H}_{0}(B,t),\mathcal{H}_{int}(B,t)]-i\partial_{t}\mathcal{H}_{int}(B,t)\ ,
\label{eq:3}
\end{equation}
Here $\mathcal{H}_{0}$ denotes the diagonal part of the flowing Hamiltonian
and $\mathcal{H}_{int}$ its interaction part, $\mathcal{H}(B,t)=\mathcal{H}_{0}(B,t)+\mathcal{H}_{int}(B,t)$.
Then one can easily show that the time-dependent interaction matrix elements satisfy an inhomogeneous
diffusion equation with a negative source term
\begin{equation}
(\partial_{B}-\partial_{t}^{2})\mathrm{Tr}\left[\mathcal{H}_{int}^{2}(B,t)\right]  =  
-2\:\mathrm{Tr}\left[\eta(B,t)\eta^\dagger(B,t)\right] \leq 0\ .
\label{eq:4}
\end{equation}
This ensures that for $B\rightarrow\infty$ only resonant
couplings $\mathcal{H}_{int}^{ij}(B\rightarrow\infty,t)=\tilde{\mathcal{H}}_{ij}e^{-i(\tilde{\mathcal{H}}_{0}^{ii}-\tilde{\mathcal{H}}_{0}^{jj})t}$ remain in the interaction part of the Hamiltonian.
Hence in general there remains a simpler time evolution problem in the $B=\infty$ basis with at most the
complexity of the common rotating wave approximation.

The main computational problem of our new approach
is that solving the set of partial differential equations (\ref{eq:2}) and (\ref{eq:3})
is more difficult
than the set of ordinary differential equations in the time-independent case. 
Notice that according to (\ref{eq:3}) the parameter~$B$ has the dimension energy${}^{-2}$,
therefore our flow occurs in an energy-scale separated way by first decoupling matrix
elements with large energy differences before looking at smaller energy differences.
This will turn out to produce an RG-like scaling flow for time-dependent Hamiltonians.

Next we illustrate this new approach for the weak coupling ferromagnetic
Kondo model (\ref{defferroKondo}), which will allow for a controlled truncation of 
the infinite set of flow equations in an expansion
in the time-dependent running coupling $J_{kk'}(B,t)$. The calculation
proceeds along similar lines to the interaction
quench in the ferromagnetic Kondo model, therefore we adopt the
notation from Refs.~\cite{hackl-2009-102,hackl-2009}.
Details of the calculation will be published separately.

\section{Time-dependent Kondo model}

To leading order in the running coupling constants (equivalent to
one loop order in the conventional scaling approach) the flowing
time-dependent Hamiltonian can be parametrized as
\begin{equation}
\mathcal{H}(B,t) = \sum_{k,\sigma} \varepsilon_{k}c_{k,\sigma}^{\dagger}c_{k,\sigma}+
\mathcal{H}_{int}(B,t)
\label{eq:5}
\end{equation}
with the interaction part
\begin{equation}
\mathcal{H}_{int}(B,t)  =  \sum_{k,k'} J_{kk'}(B,t)\,\vec S\cdot \vec s_{kk'}
\end{equation}
From (\ref{eq:2}) and (\ref{eq:3}) one derives the key result for the time-dependent
one loop scaling equation of the coupling constant: 
\begin{eqnarray}
\partial_{B}J_{kk'}(B,t) &=& -\Delta_{kk'}^{2}(t)J_{kk'}(B,t) \label{eq:7} \\
&&\hspace*{-2.0cm}
-\sum_{q}\left(J_{kq}\Delta_{qk'}(t)J_{qk'}-J_{qk'}\Delta_{kq}(t)J_{kq}\right)
\left(n_{q}-\dfrac{1}{2}\right) \nonumber
\end{eqnarray} 
where we have introduced the operator
\begin{eqnarray}
\Delta_{kk'}(t) & = & (\varepsilon_{k}-\varepsilon_{k'}-i\partial_{t}) \ .
\label{eq:6}\end{eqnarray}
Here $n_q$ is the occupation number distribution of the noninteracting conduction band.
In the sequel we will only consider the zero temperature case (the generalization
to nonzero temperature is straightforward): $n_q=\Theta(-\varepsilon_q)$ where we
have taken the Fermi energy $\varepsilon_F=0$. Notice that (\ref{eq:7}) holds
for arbitrary time-dependence and both antiferromagnetic and ferromagnetic couplings.
The neglected normal ordered terms in the flow of the Hamiltonian are third order
in the running couplings. 

For time-independent couplings Eq.~(\ref{eq:7}) reduces to the standard
one loop scaling equation of the flow equation formalism \cite{NewEntry10}.
For the time-dependent case
Eq.~(\ref{eq:7}) has the structure of a complicated non-linear diffusion equation:
An interaction that is initially localized in time spreads out during the 
$B$-evolution. Notice that this does not lead to violation of causality since all the observables
undergo the same unitary flow, therefore a time-dependent expectation value at 
time~$t$ does only depend on couplings $J(t')$ for $t'<t$.

Eq.~(\ref{eq:7}) is best analyzed in Fourier space,
\begin{equation}
J_{kk'}(\omega)=\int dt\,e^{i\omega t}\,J_{kk'}(t) \ .
\end{equation}
The solution of the linearized form of (\ref{eq:7}) gives the enveloping
behavior of the flowing couplings which we later dress with the one loop scaling flow. 
The linearized solution is simple in Fourier space: 
\begin{eqnarray}
J_{kk'}(B,\omega) & = & e^{-B\Delta_{kk'}^{2}(\omega)}J(B=0,\omega),\label{eq:8}\end{eqnarray}
with $\Delta_{kk'}(\omega)=\epsilon_k-\epsilon_{k'}-\omega$.
The initial condition $J(B=0,\omega)$ is the Fourier transform of $J(t)$
from (\ref{defferroKondo}).
Therefore our formalism with the generator (\ref{eq:3})
eliminates the non-resonant Fourier modes first in an energy scale separated way
in a direct generalization of conventional time-independent scaling theory.

Now one can parametrize the full solution of (\ref{eq:7}) very accurately (meaning: correct
in one loop order) by the following approximation
\begin{eqnarray}
J_{kk'}(B,\omega) & \equiv & e^{-B\Delta^2_{kk'}(\omega)}\,j_{\epsilon_{\overline{kk'}}}(B,\omega)
\label{eq:11}\end{eqnarray} 
where the flow of $j_{\epsilon_{\overline{kk'}}}(B,\omega)$ is determined from the flow 
of $J_{kk'}(B,\omega)$ at the point $\Delta_{kk'}(\omega)=0$. The notation
$\epsilon_{\overline{kk'}}$ corresponds to the average energy of $\epsilon_k$ and $\epsilon_{k'}$:
$\epsilon_{\overline{kk'}}\stackrel{\rm def}{=}(\epsilon_k+\epsilon_{k'})/2$. In (\ref{eq:11}) 
the flow of the dimensionless coupling $j_{\epsilon_{\overline{kk'}}}(B,\omega)$ therefore effectively
dresses the linearized solution with the one loop scaling flow.

Substituting $\Delta_{kk'}(\omega)=0$ into (\ref{eq:7}) yields the following scaling equation
for $j_\epsilon(B,\omega)$ 
\begin{eqnarray}
\frac{\partial j_\epsilon(B,\omega)}{\partial B} &=& \dfrac{1}{2B}\;\rho\int\dfrac{d{\omega'}}{2\pi}
e^{-2B(\epsilon-\omega')^{2}} \label{eq:12} \\
&&\times j_{\epsilon+\omega/2}(B,\frac{\omega}{2}+{\omega'})\, 
j_{\epsilon-\omega/2}(B,\frac{\omega}{2}-{\omega'}) \nonumber
\end{eqnarray}
Here the initial condition is posed at $B=D^{-2}$, where $D$ is the band width of 
the conduction band. $\rho$ is the density of states in the conduction band, which
we assume to be constant.

Specificially, for the time-independent case with
$J_{kk'}(B=0,t)=J_0$
one finds $j_\epsilon(B=0,\omega)=2\pi\delta(\omega)\,J_0$.
Since the flow at the Fermi surface determines the low-energy physics we use 
the identification $j_{\epsilon_F}(B,\omega)=2\pi\delta(\omega)\,J(B)$
and (\ref{eq:12}) reduces to the usual one loop scaling equation for the infrared behavior
\begin{equation}
\frac{dJ}{dB}=\frac{1}{2B}\,\rho\, J^2
\end{equation}
or in terms of the scaling parameter $\Lambda=B^{-1/2}$ with dimension energy:
\begin{equation}
\frac{dJ}{d\ln\Lambda}=-\rho\, J^2
\end{equation}
The solution shows the characteristic logarithmic scaling of the Kondo model,
\begin{equation}
J(B)=\frac{J_0}{1-\frac{\rho\,J_0}{2}\ln(B\,D^2)} \ .
\label{eq:staticJ}
\end{equation}

\section{Continuous switching of the exchange coupling}

In our model we are interested in a situation where the exchange interaction $J(t)$
is switched on continuously over a timescale $\tau$. For convenience
we choose the specific form
\begin{eqnarray}
J_{kk'}(B=0,t) & = & \dfrac{J_0}{2}\:\left(1+\mathrm{erf}\left(\dfrac{t}{2\tau}\right)\right)
\label{eq:9}\end{eqnarray}
with ferromagnetic $J_0<0$: $\lim_{t\rightarrow -\infty} J(t)=0$ and $\lim_{t\rightarrow +\infty} J(t)=J_0$.
The Fourier transform of (\ref{eq:9}) is 
\begin{equation}
J_{kk'}(B=0,\omega)= J_0\:\frac{ie^{-\tau^2\omega^2}}{\omega+i0^+} \ ,
\label{eqinitialcond}
\end{equation}
which permits simple analytic expressions in our subsequent calculations. 
For example the linearized solution (\ref{eq:8})
of (\ref{eq:7}) given the initial condition (\ref{eqinitialcond}) takes the following form
\begin{eqnarray}
J_{kk'}(B,t) &=& \dfrac{J_0}{2}\,e^{-B(\epsilon_k-\epsilon_{k'})^{2}} \label{eq:10} \\
&&\times\left(1 +  \mathrm{erf}\left(\dfrac{-iB(\epsilon_k-\epsilon_{k'})+t/2}{\sqrt{B+\tau^{2}}}\right)\right)
\nonumber
\end{eqnarray} 
Notice however that the formalism can be carried through for an arbitrary switching protocol. 

Next we need to calculate how the one loop scaling flow (\ref{eq:12}) dresses the linearized solution.
Using the Dirac identity one can verify that the parametrization
\begin{equation}
j_{\epsilon_F}(B,\omega)=g(B,\omega)\:\frac{ie^{-\tau^2\omega^2}}{\omega+i0^+} 
\end{equation}
in (\ref{eq:12}) leads to the following scaling equation
\begin{equation}
\frac{\partial g(B,\omega)}{\partial B}= \frac{1}{2B}\:\rho\,g(B,0)\,g(B,\omega)\:
e^{-B\omega^2/2} \ ,
\label{eq:14}\end{equation} 
with the initial condition $g(B=D^{-2},\omega)=J_0$.
One can verify that the shift $\epsilon_F\pm\omega/2$ on the rhs of (\ref{eq:12})
can be neglected in one loop order in the derivation of (\ref{eq:14}).
Eq.~(\ref{eq:14}) can be solved easily: For $B\lesssim \omega^{-2}$ the flow is to a very good approximation 
(meaning: correct in one loop order) just the static flow (\ref{eq:staticJ})
\begin{equation}
g(B,\omega)=\frac{J_0}{1-\frac{\rho J_0}{2}\ln(B\,D^2)}
\label{eq:gflow}
\end{equation}
which stops once the energy scale $B\approx\omega^{-2}$ is reached,
\begin{equation}
g(B\gtrsim \omega^{-2},\omega)=\frac{J_0}{1-\frac{\rho J_0}{2}\ln(D^2/\omega^{2})} \ .
\label{eq:gfinal}
\end{equation}
Remarkably the switching scale $\tau$ does not enter in these equations. Notice
that we find the expected weak coupling flow for the ferromagnetic case, which
makes our calculation controlled by a small parameter.

Summing up we have derived the following one loop flow of the coupling constants
\begin{equation}
J_{kk'}(B,\omega)= g(B,\omega)\:\frac{i\,e^{-\tau^2\omega^2}}{\omega+i0^+}\:e^{-B\Delta^2_{kk'}(\omega)}
\label{eq:flowingcoupling}
\end{equation}
with $g(B,\omega)$ according to (\ref{eq:gflow}) and (\ref{eq:gfinal})
for energies around the Fermi surface, $(\epsilon_k+\epsilon_{k'})/2=\epsilon_F$. 

\section{Evaluating observables}

Next we want to evaluate observables making use of the diagonal structure of
the Hamiltonian for $B=\infty$. Specifically, let us consider an observable~$O$
without explicit time dependence and an initial state~$|\Psi_I(t_0)\rangle$ given at time~$t_0$.
Then the expectation value of~$O$ at time~$t$ is
\begin{equation}
\langle O(t)\rangle_I \stackrel{\rm def}{=}\langle\Psi_I(t)|O|\Psi_I(t)\rangle
\end{equation}
where $\Psi_I(t)$ follows from the solution of the time-dependent Schr\"odinger equation
with respect to~$H(t)$. We can rewrite this in the Heisenberg picture
\begin{equation}
\langle O(t)\rangle_I=\langle\Psi_I(t_0)|O(t,t_0)|\Psi_I(t_0)\rangle
\label{eq:defOtt0}
\end{equation}
with
\begin{equation}
O(t,t_0)= U^\dagger(t_0)\,P^\dagger(t,t_0)\,U(t)\,O\,U^\dagger(t)\,P(t,t_0)\,U(t_0)
\label{eq:Heisenberg}
\end{equation}
and the time evolution $P(t,t_0)$ evaluated in the simple $B=\infty$ basis:
\begin{equation}
P(t,t_0)\stackrel{\rm def}{=} {\cal T}_t\left[ \exp\left(-i\int_{t_0}^{t} dt'\,{\cal H}(t')\right)\right]
\end{equation}
$O(t,t_0)$ from (\ref{eq:Heisenberg}) is therefore the solution of the equations
of motion for the operator~$O$ with the initial condition $O(t=t_0,t_0)=O$.

The explicit evaluation of (\ref{eq:Heisenberg}) proceeds similar to the 
forward-backward flow equation transformation scheme for time-independent 
Hamiltonians \cite{Hackl2008}. First we derive 
$\tilde O(t)=U(t)\,O\,U^\dagger(t)$ by solving the differential flow equation
\begin{equation}
\frac{O(B,t)}{dB}=[\eta(B,t),O(B,t)]
\label{eq:flowOBt}
\end{equation}
with the initial condition $O(B=0,t)=O$ and the result $\tilde O(t)=O(B=\infty)$.
Next we employ the time evolution
\begin{equation}
\tilde O(t,t_0) \stackrel{\rm def}{=} P^\dagger(t,t_0)\,\tilde O(t)\,P(t,t_0)
\label{eq:P}
\end{equation}
which will usually be simple since ${\cal H}$ is diagonal. Then we undo the transformation
and go back into the original basis, $O(t,t_0)=U^\dagger(t_0)\,\tilde O(t,t_0)\,U(t_0)$ by solving
the differential equation
\begin{equation}
\frac{O(B,t,t_0)}{dB}=[\eta(B,t_0),O(B,t,t_0)]
\label{eq:trfback}
\end{equation}
with the initial condition $O(B=\infty,t,t_0)=\tilde O(t,t_0)$ now posed at $B=\infty$. 
The solution at $B=0$ gives the desired time evolved operator for (\ref{eq:defOtt0}), 
$O(t,t_0)=O(B=0,t,t_0)$. 
The main difference to the time-independent case \cite{Hackl2008} is that here
the forward and backward unitary transformations $U(t)$ and $U(t_0)$ will generically be different. 
The key advantage of our approach is like in Ref.~\cite{Hackl2008}
that there are no secular terms in time (i.e. errors that grow with
$t-t_0$) since we can evaluate (\ref{eq:P}) exactly for diagonal Hamiltonians in the $B=\infty$ basis.

\section{Time evolution of the impurity spin}

We want to use the above framework to derive the time evolution of the impurity spin 
expectation value starting from an initial product state with a fully polarized spin
at $t_0=-\infty$:
\begin{equation}
|\Psi_I\rangle = |{\rm FS}\rangle \,\otimes\, |\uparrow\rangle
\end{equation}
where $|{\rm FS}\rangle$ denotes the zero temperature Fermi sea. For the unitary
transformation of the observable $O=S_z$ we make the following ansatz \cite{NewEntry10,hackl-2009}
\begin{equation}
S_{z}(B,t)  =  h(B,t)\,S_{z}+
\sum_{k,k'}\gamma_{kk'}(B,t)\,(S^{+}s_{k'k}^{-}-S^{-}s^+_{k'k})
\label{eq:19}\end{equation}
where we neglect higher order terms in order $O(J^3)$. 
The flow equations (\ref{eq:flowOBt})
for the coefficients are structurally identical to Refs.~\cite{NewEntry10,hackl-2009}
with an additional time dependence:
\begin{eqnarray}
\hspace{-0.5em}\partial_{B}\gamma_{kk'}(B,t) & \hspace{-0.5em}= & \hspace{-0.5em}h(B,t)\Delta_{kk'}(t)J_{kk'}(B,t) \label{eq:20}\\
\hspace{-0.5em}\partial_{B}h(B,t) & \hspace{-0.5em}=\hspace{-0.5em} & 
\sum_{k,k'}\gamma_{k'k}(B,t)\Delta_{kk'}(t)J_{kk'}(B,t)n_{k}(1-n_{k'})\nonumber 
\end{eqnarray}
It is possible to verify that one can use the solution (\ref{eq:flowingcoupling})
for all values $k,k'$ in these equations, the deviation only plays a role in higher orders.
The solution for the Fourier transform at $B=\infty$ with the initial condition
$h(B=0,t)=1$, $\gamma_{kk'}(t)=0$ is readily obtained as
\begin{equation}
\gamma_{kk'}(B=\infty,\omega)=g\left((\epsilon_k-\epsilon_{k'})^{-2},0\right)\:
\frac{i\,e^{-\tau^2\omega^2}}{\Delta_{kk'}(\omega)\,(\omega+i0)}
\end{equation}
Time evolution (\ref{eq:P}) just generates additional phase factors
$e^{i(\epsilon_k-\epsilon_{k'})(t-t_0)}$.
Inserting the result back in (\ref{eq:trfback}) and integrating back to $B=0$ yields
\begin{eqnarray}
\gamma_{kk'}(t,t_0)&=&g\left((\epsilon_k-\epsilon_{k'})^{-2},\omega=0\right)
\label{eq:gamma} \\
&&\times \int\frac{d\omega}{2\pi}\,
\frac{i\,e^{-\tau^2\omega^2}}{\Delta_{kk'}(\omega)\,(\omega+i0)} \nonumber \\
&&\times \left( e^{i(\epsilon_k-\epsilon_{k'})(t-t_0)}\,e^{-i\omega t}
-e^{-i\omega t_0}\right) \nonumber
\end{eqnarray}
and
\begin{equation}
h(t,t_0)=1+\frac{1}{2}\sum_{k.k'} \gamma_{kk'}(t,t_0)\, \gamma_{k'k}(t,t_0)\,
n_k(1-n_{k'})
\label{eq:h}
\end{equation}
plus terms that are third order in the running coupling. Eqs.~(\ref{eq:gamma})
and (\ref{eq:h}) therefore constitute the solution for the operator equation
of motion in the original basis and we can immediately read off the desired
spin expectation value 
\begin{equation}
\langle S_z(t)\rangle_I=\lim_{t_0\rightarrow -\infty} \langle \Psi_I|S_z(t,t_0)|\Psi_I\rangle
=\frac{1}{2}\, h(t,t_0\rightarrow -\infty)
\end{equation}
Now one can easily work out
\begin{eqnarray}
\gamma_{kk'}(t,t_0)&=&g\left((\epsilon_k-\epsilon_{k'})^{-2},\omega=0\right)\,
\frac{e^{-i(\epsilon_k-\epsilon_{k'})t_0}}{\epsilon_k-\epsilon_{k'}} \nonumber \\
&&\times \: \left(e^{i(\epsilon_k-\epsilon_{k'})t} - e^{-\tau^2(\epsilon_k-\epsilon_{k'})^2}\right)
\end{eqnarray}
for $t\gg \max(\tau,D^{-1})$, $t_0 \ll -\max(\tau,D^{-1})$ which yields
\begin{eqnarray}
\langle S_{z}(t)\rangle_I & = & \frac{1}{2}-\frac{1}{4}\int_0^D\,d\epsilon\,
\frac{\rho^2 J^2(\epsilon)}{\epsilon} \nonumber \\
&& \times\left( 1 + e^{-2\tau^2\epsilon^2} -2e^{-\tau^2\epsilon^2}\cos(\epsilon t)\right)
\label{eq:28}
\end{eqnarray}
where $J(\epsilon)=J_0/\left(1-\rho J_0\ln(D/\epsilon)\right)$ is just the usual infrared
running coupling constant (\ref{eq:staticJ}) of the time-independent Kondo model.
Eq.~(\ref{eq:28}) is our key result which contains both the asymptotic
magnetization for $t\rightarrow\infty$ and the approach to it as a function
of the switching time scale~$\tau$. It can be viewed as a systematic resummation
of the short time perturbative expansion \cite{Schiller} that incorporates
both scaling effects and the switching protocol (\ref{eq:9}). Notice that while
we have made the simplifying assumption $t\gg \tau$ in deriving (\ref{eq:28}),
one can obtain the behavior on all time scales from (\ref{eq:gamma}) and (\ref{eq:h}).

\begin{figure}
\begin{center}
\includegraphics{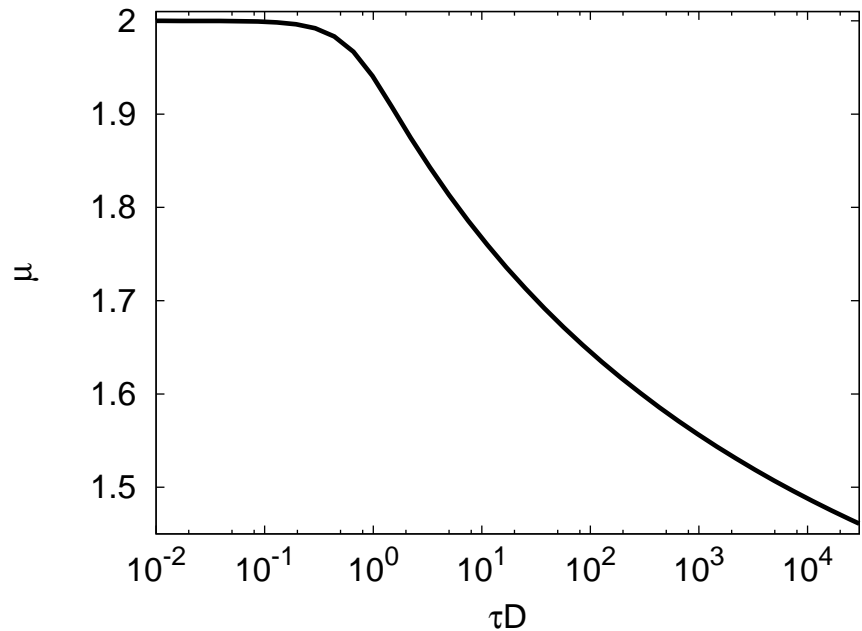}
\par \caption{Mismatch factor $\mu(\tau)$ from (\ref{eq:mu}) for $\rho J_0=-0.1$.}
\label{figure:mu}
\end{center}
\end{figure}

Eq.~(\ref{eq:28}) shows the correct limiting behavior for adiabatic switching on
($\tau\rightarrow\infty$)
\begin{equation}
\langle S_{z}(t=\infty)\rangle_I=\frac{1}{2}-\frac{1}{4}\int_0^D\,d\epsilon\,
\frac{\rho^2 J^2(\epsilon)}{\epsilon}=\frac{1}{2}+\frac{\rho J_0}{4}+O(J_0^2)
\label{Szeq}
\end{equation}
which reproduces the equilibrium magnetization for an infinitesimal magnetic 
field \cite{Abrikosov}. For instantaneous switching $\tau=0$ Eq.~(\ref{eq:28})
reproduces the results from Refs.~\cite{hackl-2009-102,hackl-2009}
\begin{equation}
\langle S_{z}(t=\infty)\rangle_I=\frac{1}{2}+\frac{\rho J_0}{2}+O(J_0^2)
\label{Szneq}
\end{equation}
The mismatch factor $\mu$ introduced in (\ref{defmuSz}) interpolates
between (\ref{Szeq}) and (\ref{Szneq}) as a function of the switching time~$\tau$,
\begin{equation}
\langle S_{z}(t=\infty)\rangle_I=\frac{1}{2}+\mu(\tau)\:\frac{\rho J_0}{4}+O(J_0^2)
\end{equation}
and can be derived from (\ref{eq:28})
\begin{equation}
\mu(\tau)=\int_0^D\: d\epsilon\: \left(1+e^{-2\tau^2\epsilon^2}\right)\:
\frac{\partial}{\partial\epsilon} \left(\frac{1}{1+\rho J_0\,\ln(\epsilon/D)}\right)
\label{eq:mu}
\end{equation}
It is depicted in Fig.~\ref{figure:mu} and shows the expected crossover from~2
(instantaneous quench) to~1 (adiabatic quench equivalent to equilibrium). Notice 
that~$\mu(\tau)$ starts to deviate from~2 when $\tau$ becomes of order the inverse
bandwidth since then there are fast modes with $\epsilon \gtrsim \tau^{-1}$
for which the quench is no longer instantaneous. However, the approach to the
equilibrium value~1 is very slow, 
\begin{equation}
\mu(\tau\gg D^{-1}) =1+\frac{1}{1-\rho J_0\,\ln(\tau D)}
\end{equation}
For example the value $\mu(\tau)=3/2$ is only reached at the nonperturbative time
scale $\tau=D^{-1}\,\exp\left(-\,\frac{1}{|\rho J_0|}\right)$.

Likewise the approach of the spin expectation value to its asymptotic value
is also logarithmic and (remarkably) independent of~$\tau$ as can be derived
from (\ref{eq:28})
\begin{equation}
\langle S_{z}(t)\rangle_I = \langle S_{z}(t=\infty)\rangle_I 
-\frac{1}{2}\,\frac{\rho\,J_0}{1-\rho J_0\,\ln(t\,D)}
\end{equation}
for $t \gg \max(\tau,D^{-1})$.

\section{Conclusion}

We illustrated how a generalization of the flow equation
method can be used to study time-dependent Hamiltonians with a nontrivial scaling behavior. 
In particular for the time-dependent ferromagnetic Kondo model we calculated the mismatch
factor~$\mu$ which measures the ``non-adiabacity'' of the quench. Different from previously
studied models like the quench in the Hubbard model \cite{moeckel-2008-100,moeckel3,moeckel-2009} 
here the approach of $\mu$ to the equilibrium value is only logarithmic and one needs
to quench very slowly to reach the adiabatic limit. 

\section*{Acknowledgment}
This work was supported
through SFB/TR~12 of the Deutsche Forschungsgemeinschaft, the Center for NanoScience (CeNS) Munich, and the German Excellence Initiative via the Nanosystems Initiative Munich (NIM).

\end{document}